\documentclass[12pt,noshowpacs,nofootinbib,notitlepage,amsmath]{elsarticle}
\usepackage{setspace}
\usepackage{amsmath,amsfonts}
\usepackage{graphicx,color}
\usepackage{enumerate}
\usepackage{array}
\usepackage{bm}

\begin{document}

\title{Making sense of ghosts}
\author{Bob Holdom}
\ead{bob.holdom@utoronto.ca}
\affiliation{Department of Physics, University of Toronto, Toronto, Ontario, Canada  M5S 1A7}
\begin{abstract}
Ghosts have been a stumbling block in the development of a UV complete quantum field theory for gravity. We discuss how difficulties associated with ghosts are overcome in the context of 0+1d QFT. Obtaining a probability interpretation is the key issue, and for this we discuss how an appropriate inner product can be constructed to define a sensible Born rule. Ghost theories are intrinsically unitary and perturbatively stable. They can also display nonperburbative stability even when the corresponding normal theory does not. The spectra and propagators are numerically obtained at both weak and strong coupling. Normalizable wave functions are obtained for the energy eigenstates and they show a violation of normal parity. We discuss connections to PT-symmetric quantum mechanics.
\end{abstract}
\maketitle

\section{Introduction}

A renormalizable quantum field theory of gravity \cite{stelle} has a massive spin-2 partner of the graviton. This is a ghost, a field with a wrong sign kinetic term. The quantization that gives rise to renormalizability also ensures that all perturbative states have positive energy. But this quantization implies that the ghost state has negative norm. The problem with negative norms is not with S-matrix unitarity. This is still satisfied as $S^\dagger\textbf{1}S=\textbf{1}$, where a generalized identity operator reflects the negative norms via the completeness relation $\textbf{1}=\sum_X\frac{|X\rangle \langle X|}{\langle X|X \rangle}$. Equivalently the optical theorem can be seen to be satisfied at the diagrammatic level when properly accounting for both the wrong sign of the ghost propagator and the negative norm of the ghost state.

The issue is that we are no longer guaranteed that probabilities, and quantities such as cross sections, are positive. A quantum mechanical measurement that causes the initial and final states to differ in the sign of their norms will produce a negative probability via the Born rule $P=|\langle f|i\rangle|^2/(\langle f|f\rangle\langle i|i\rangle)$, and a probability interpretation is lost. A way forward comes when noting that the formulation of the Born rule to extract probabilities is an aspect of quantum theory that is quite separate from unitarity. In fact the ``add-on'' nature of the Born rule among the postulates of quantum mechanics is wrapped up with the measurement problem. Closer to the core of quantum mechanics is unitarity, and it is the unitary dynamics of a quantum theory that determines its spectrum, as well as the correlation functions of interest in QFT. This suggests that some modification of the Born rule may be made independently of all that unitarity entails. The question then arises: can a sensible Born rule be achieved and is it unique?

We can also mention a counter-example to the belief that negative norms necessarily imply negative probabilities, even without modifying the Born rule. Consider a ghost theory that has a conserved ghost parity, where the ghost parity of a state is the sign of its norm. In such a theory the ghost parity of initial and final states must be the same, and so probabilities are always positive. Thus ghost theories with ghost parity have a probability interpretation even with wrong sign propagators and negative norms.

Developing a QFT for gravity requires us to consider theories that do not conserve ghost parity. In this case a probability interpretation will require a modified Born rule, and this shall be implemented via an inner product to be used exclusively in the formulation of the Born rule. This new inner product differs from the original inner product that emerges in the construction of the theory, and which is already preserved under time evolution. The new inner product must also be preserved under time evolution, but it must yield all positive norms as well. It is still convenient to speak of states with negative norms with respect to the original inner product, and these states continue to play their role in unitary evolution. But when probabilities involving these states are calculated with the new Born rule, the probabilities are positive.

In quantum field theory, positivity constraints can be placed on various quantities when they can be related to the positivity of probabilities. One such quantity is the spectral function of the propagator. Such constraints have an implicit dependence on the standard Born rule. If a new Born rule disrupts the link between negative norms and negative probabilities then this disrupts the positivity proof for the spectral function, since the proof relies on this link to rule out negative norm states. Similarly we have mentioned that new signs are introduced into the optical theorem due to negative norms. Arguments about positivity that rely on the optical theorem are no longer valid, since a reformulated Born rule means that there is no longer a link between unusual signs in the optical theorem and negative probabilities. For the gravity QFT we note in particular that integrating out the massive spin-2 ghost will produce low energy amplitudes with signs not agreeing with the standard arguments \cite{Holdom:2021oii}.

The problem of a QFT for gravity has already spurred interest in two extensions of quantum mechanics where modified inner products are considered. One is PT-symmetric quantum mechanics, as discussed in this context by Mannheim and Bender \cite{BM1,BM2,Mannheim:2009zj}, and the other is Dirac-Pauli quantization, as discussed in the same context by Salvio and Strumia \cite{Salvio:2015gsi,Strumia:2017dvt,Salvio:2018crh}. We shall start with a canonical quantization that is tailored to a ghost theory and see where it leads. The path we follow will sometimes have overlap with one or the other of these other approaches.

We shall also explore how a positive energy spectrum emerges in an interacting ghost theory, and how these theories enable new dynamics that have no analog among normal theories. For example a normal $\phi^3$ theory develops an energy spectrum that is not bounded from below. In contrast we find that a cubic ghost theory has a non-negative energy spectrum, for both weak and strong coupling, that is not enjoyed by its normal theory partner. This goes against the intuition derived from classical theories, where either a negative kinetic energy term or a cubic interaction are both separately associated with instability.

The quantum mechanical ghost models that we study are based on Hermitian Hamiltonians, which ensures unitary evolution. The quantization ensures that the free energy spectrum is positive even though the quadratic part of the Hamiltonian has the wrong sign. To determine the spectrum at both weak and strong coupling we develop a matrix notation for the study of the eigenvalue problem for the full Hamiltonian. The negative norms have the effect that the full matrix Hamiltonian that enters the eigenvalue problem may be non-Hermitian. The problem of diagonalizing an infinite matrix is approached by truncating to finite size matrices and by studying the behavior of the resulting spectra as the matrix size is increased. The spectrum of the interacting ghost theory converges to a positive one in the infinite size limit. In the context of 0+1d QFT we also obtain the full propagator from the energy eigenvalues and eigenstates.

PT-symmetric quantum mechanics \cite{Bender:2023cem} is concerned with non-Hermitian Hamiltonians. These theories are also able to yield real and positive spectra, and we shall find that certain PT-symmetric theories have spectra that match the spectra of corresponding ghost theories. But unlike ghost theories, an inner product consistent with unitarity does not emerge automatically in PT-symmetric theories. Also, typically, a complex extension of the coordinate-space representation is used, with the result being that the position and momentum operators do not correspond to observables. We shall present position and momentum operators for ghost theories that do correspond to observables. The resulting eigenvalues, the positions and the momenta, are real and the resulting wave functions are normalizable. Wave functions for the energy eigenstates are obtained and studied as a function of coupling.

Ghost theories with a finite dimensional truncation of their Hilbert space, as mentioned above, are of interest to study for their own sake. In these theories, for sufficiently large coupling, there exist pairs of states with complex-conjugate energy eigenvalues. Here we find that if we restrict to the space of states contributing to the spectral representation of the propagator, then we are again able to find a conserved inner product and positive norms. In 3+1d QFTs, pairs of poles at complex-conjugate positions on the complex plane can arise due to perturbative corrections to propagators. This seems especially problematic in a ghost theory, but our findings suggest that a probability interpretation may be possible even in this case.

In the final section of this paper we provide numerical results for some extensions of the single field model, again for weak and strong coupling. We start with a model of a ghost interacting with a normal field, where we find that a positive spectrum survives the interactions. This is perhaps not surprising, given that the perturbative energies of both the ghost and the normal field are positive. We then go on to introduce more than one spatial point and the discrete analog of the spatial derivative terms. These are rudimentary models of 1+1d QFTs with ghosts. This gives a first indication of how the dynamics of ghost 0+1d QFT can extend to higher dimensional QFTs.

\section{Preliminaries}

We start with a Hamiltonian having an interaction term with coupling $\lambda$ and a sign $\sigma$ to control whether the quadratic terms are normal $\sigma=1$ or ghostly $\sigma=-1$,
\begin{align}
    H_{\sigma} & = \frac{\sigma}{2}(\pi^2 + m^2\phi^2) + \frac{\lambda}{k!}\phi^k\quad\textrm{with}\quad [\phi,\pi] = i
.\label{e1}\end{align}
Making the replacements
\begin{align}
    \phi = \frac{1}{\sqrt{2m}}(a + a^\dagger),\quad
    \pi = \frac{\sqrt{m}}{i\sqrt{2}}(a - a^\dagger),
\label{e5}\end{align}
gives
\begin{align}
    H_{\sigma} & = \frac{\sigma}{2} m\left( a^\dagger a + a a^\dagger\right) + \frac{\lambda}{(2m)^\frac{k}{2}k!}(a+a^\dagger)^k\quad\textrm{with}\quad [a,a^\dagger] =1
.\end{align}
Other than the sign $\sigma$ in the Hamiltonian, everything here is standard. The Hamiltonian is Hermitian with the dimensionful constants $m$ and $\lambda$ being real, and $k$ is an integer $k\geq 3$. We shall see how the first term in (\ref{e1}) corresponds to a positive free spectrum for both $\sigma=\pm 1$, and we will have more to say about the sign of $\lambda$.

We introduce a vacuum state $|0\rangle_\sigma$ with unit norm,
\begin{align}
    _{\sigma}\langle 0 | 0 \rangle_\sigma  = 1
.\end{align}
The Hilbert space is spanned by the occupation number basis, and this is constructed depending on $\sigma$,
\begin{align}
    a |0\rangle_+  &= 0, \quad\;\; | n \rangle_+ = \frac{1}{\sqrt{n!}} (a^\dagger)^n |0\rangle_+,\quad n\geq 1,\label{e6}\\
    a^\dagger |0\rangle_-  &= 0, \quad\;\; | n \rangle_- = \frac{1}{\sqrt{n!}} a^n |0\rangle_-,\quad n\geq 1
.\label{e21}\end{align}
We see that the construction for the ghost theory in (\ref{e21}) has $a$ and $a^\dagger$ interchanged relative to a normal theory in (\ref{e6}).\footnote{The basic setup for ghost quantization in \cite{Salvio:2015gsi,Strumia:2017dvt,Salvio:2018crh} looks different but is equivalent.} Powers of $\sigma$ then show up in the norms, which alternate in sign for $\sigma=-1$,
\begin{equation}
    {_\sigma}\langle m | n \rangle_\sigma  = \sigma^n\delta_{mn}.
\end{equation}
These powers of $\sigma$ then also appear in the completeness relation,
\begin{equation}
    \textbf{1} = \sum_{n \geq 0} \frac{| n \rangle_\sigma {_\sigma}\langle n |}{{_\sigma}\langle n | n \rangle_\sigma} = \sum_{n\geq 0} \big(| n \rangle_\sigma {_\sigma}\langle n | \big) \sigma^n,
\end{equation}
and when solving for the coefficients of a general state,
\begin{align}
    |\psi\rangle_\sigma &= \sum_{n\geq 0} \psi_{n,\sigma} |n \rangle_\sigma,\quad\quad\psi_{n,\sigma} = ( {_\sigma}\langle n | \psi \rangle_\sigma)\sigma^n
.\end{align}

We find that a matrix notation helps to clarify the meaning of ghost quantization. We introduce a bold notation for an infinite matrix $\textbf{A}$ and an infinite column vector $\bm\psi$, with components
\begin{align}
    (\textbf{A}_\sigma)_{mn} & = {_\sigma}\langle m | A | n \rangle_\sigma, \\
    (\bm{\psi}_\sigma)_n &=\psi_{n,\sigma}.
\end{align}
We also introduce the matrix $ \bm{\eta}_\sigma = \textrm{diag}(\sigma^0,\sigma^1,\sigma^2,\cdots)$ so that $\bm{\eta}_-$ represents a ghost parity operator in the occupation number basis. Note that in general the matrix representation of an operator depends on $\sigma$. In particular the results in (\ref{e5}-\ref{e21}) imply that
\begin{align}
     \bm{\phi}_\sigma=\begin{bmatrix}
    0      & \sigma       & 0        & 0 & \cdots \\
    \sigma     & 0        & \sqrt{2}  & 0 & \cdots \\
    0      & \sqrt{2} & 0         & \sigma\sqrt{3} & \cdots\\ 
    0 & 0        & \sigma\sqrt{3} & 0 & \\
    \vdots & \vdots        & \vdots &  & \ddots
    \end{bmatrix},\quad\quad
        \bm{\pi}_\sigma=i\begin{bmatrix}
    0      & -1       & 0        & 0 & \cdots \\
    1     & 0        & -\sigma\sqrt{2}  & 0 & \cdots \\
    0      & \sigma\sqrt{2} & 0         & -\sqrt{3} & \cdots\\ 
    0 & 0        & \sqrt{3} & 0 & \\
    \vdots & \vdots        & \vdots &  & \ddots
    \end{bmatrix}
.\label{e28}\end{align}

Now that the various dependencies on $\sigma$ are clear, we shall drop the $\sigma$ subscript henceforth. Our results will still encompass both cases $\sigma=\pm1$, but for the most part we discuss the ghost theory. The translation of an operator product $AB$ to matrix notation can be obtained by inserting a complete set of states, which yields $\textbf A\bm\eta\textbf B$. Similarly, the matrix form of the inner product is
\begin{align}
    \langle \psi|\chi\rangle= \bm\psi^\dagger\bm\eta\bm\chi =(\bm\eta\bm\psi)^\dagger\bm\chi
.\label{e37}\end{align}

An operator $A$ is self-adjoint if $\langle\psi|A\chi\rangle=\langle A\psi|\chi\rangle$, and this translates to
\begin{align}
    \textbf{A}^\dagger\bm\eta=\bm\eta\textbf{A}
.\end{align}
We shall say that an operator $A$ is Hermitian if $\textbf{A}^\dagger=\textbf{A}$. If we introduce the notation
$$\tilde{\textbf{A}}\equiv\bm\eta\textbf{A},$$ then we have the two statements ($\textbf{A}$ is self-adjoint iff $\tilde{\textbf{A}}$ is Hermitian) and ($\textbf{A}$ is Hermitian iff $\tilde{\textbf{A}}$ is self-adjoint).  The ghost matrix Hamiltonian is Hermitian, $\textbf{H}^\dagger=\textbf{H}$, and so the self-adjoint matrix $\tilde{\textbf{H}}\equiv\bm\eta\textbf{H}$ shall play a special role.

\section{The matrix Hamiltonian}

From the unitary evolution of an arbitrary state
\begin{equation}
    |\psi(t)\rangle_\sigma = \exp(-i H_{\sigma} t) | \psi \rangle_\sigma,
\end{equation}
we have
\begin{equation}
    \langle m | \psi(t) \rangle = \sum_{n\geq 0} \langle m | \exp(-i t H) | n \rangle \sigma^n \langle n | \psi \rangle 
.\end{equation}
Expressing this in matrix notation and expanding gives
\begin{align}
    \bm{\eta} \bm{\psi}(t)
    & = \left[\bm{\eta} + (-i t) \textbf{H} + \frac{(-i t)^2}{2!} \textbf{H}\bm{\eta} \textbf{H} + \cdots \right]\bm{\psi}, \\
  \bm{\psi}(t)  & =
    \left[I + (-i t) \bm{\eta} \textbf{H} + \frac{(-i t)^2}{2!} (\bm{\eta} \textbf{H})^2 + \cdots \right]\bm{\psi}.
\end{align}
Thus it is the matrix $\tilde{\textbf{H}}=\bm\eta\textbf{H}$ that describes the matrix form of time evolution,
\begin{align}
   \bm{\psi}(t) = \exp(-i t \tilde{\textbf{H}}) \bm{\psi}
.\label{e17}\end{align}
Since $\tilde{\textbf{H}}$ is self-adjoint, that is $\tilde{\textbf{H}}^\dagger\bm\eta=\bm\eta\tilde{\textbf{H}}$, we also have
\begin{align}
    (\bm\eta\bm\psi(t))^\dagger=(\bm\eta\bm\psi)^\dagger \exp(i t \tilde{\textbf{H}}).
\end{align}
We therefore see that the time dependence cancels in $\langle \psi(t) | \chi(t) \rangle=(\bm\eta\bm\psi)^\dagger\bm\chi$ and so the inner product is preserved under time evolution. Note that while the Dirac inner product $\bm\psi^\dagger\bm\chi$ produces only positive norms, it is not preserved under time evolution and so it is not tenable.

With a preserved inner product we may consider the Born rule. This should give a probability that a measurement causes a transition from some initial state $\bm\psi_i$ to some final state $\bm\psi_f$. We try
\begin{align}
    \textrm{Pr}(i\to f)=\frac{|(\bm\eta\bm\psi_f)^\dagger\bm\psi_i|^2}{((\bm\eta\bm\psi_i)^\dagger\bm\psi_i)((\bm\eta\bm\psi_f)^\dagger\bm\psi_f)}
.\label{e34}\end{align}
The result $\sum_f\textrm{Pr}(i\to f)=1$ is automatic and it is related to both unitarity and completeness. But when ghost parity is not preserved then (\ref{e34}) does not ensure that $0\leq\textrm{Pr}(i\to f)\leq1$ for all $i$ and $f$. We shall return in Section \ref{s5} to the question of whether the theory supports another inner product that produces a sensible Born rule.

Let the energies $E_n$ and the states $|\bar n\rangle$ be the eigenvalues and eigenstates of the full Hamiltonian,
\begin{align}
    H |\bar n\rangle  = E_n |\bar n \rangle,\quad n=0,1,2,\dots
.\end{align}
We denote the column vectors corresponding to $|\bar n\rangle$ by $\bm\psi^{(n)}$. The matrix equation derived in a similar way to (\ref{e17}) is
\begin{equation}
    \tilde{\textbf{H}} \bm{\psi}^{(n)} =E_n \bm{\psi}^{(n)}
.\label{e4}\end{equation}
The $E_n$ need not be real when $\tilde{\textbf{H}}^\dagger\neq\tilde{\textbf{H}}$. We do have $\tilde{\textbf{H}}^\dagger\bm\eta=\bm\eta\tilde{\textbf{H}}$ and so
\begin{equation}
   (\bm\eta\bm\psi^{(n)})^\dagger \tilde{\textbf{H}}  =E_n^* (\bm\eta\bm\psi^{(n)})^\dagger
.\label{e29}\end{equation}
(\ref{e4}) and (\ref{e29}) are describing two types of energy eigenvectors of $\tilde{\textbf{H}}$, the right- and left-eigenvectors respectively.  By multiplying (\ref{e29}) on the right by $\bm{\psi}^{(n)}$ and using  (\ref{e4}) we see that $E_n$ must be real except when the norm $(\bm\eta\bm\psi^{(n)})^\dagger  \bm\psi^{(n)}$ vanishes. When the norm vanishes there is a pair of eigenvectors with complex-conjugate energy eigenvalues. 

Let us study the Hamiltonian more closely to determine if and when complex energies occur. The terms in $\tilde{\textbf{H}}$ can be expressed as products of tilde-matrices. Specifically $\tilde{\textbf{H}}$ can be directly obtained from $H$ by replacing $\pi\to\sqrt{m/2}\;\tilde{\bm{\pi}}$ and $\phi\to\sqrt{1/2m}\;\tilde{\bm{\phi}}$ with
\begin{align}
    \tilde{\bm{\phi}}=\bm\eta\bm{\phi}=\begin{bmatrix}
    0      & \sigma       & 0        & 0 & \cdots \\
    1     & 0        & \sigma\sqrt{2}  & 0 & \cdots \\
    0      & \sqrt{2} & 0         & \sigma\sqrt{3} & \cdots\\ 
    0 & 0        & \sqrt{3} & 0 & \\
    \vdots & \vdots        & \vdots &  & \ddots
    \end{bmatrix},\quad\quad
        \tilde{\bm{\pi}}=\bm\eta\bm{\pi}=i\begin{bmatrix}
    0      & -1       & 0        & 0 & \cdots \\
    \sigma     & 0        & -\sqrt{2}  & 0 & \cdots \\
    0      & \sigma\sqrt{2} & 0         & -\sqrt{3} & \cdots\\ 
    0 & 0        & \sigma\sqrt{3} & 0 & \\
    \vdots & \vdots        & \vdots &  & \ddots
    \end{bmatrix}
.\label{e32}\end{align}
These matrices are not Hermitian when $\sigma=-1$. The quadratic (free) part of $\tilde{\textbf{H}}$ is simple,
\begin{align}
    \frac{\sigma}{2}(\tilde{\bm{\pi}}^2 + m^2\tilde{\bm{\phi}}^2)=m\bm h,\quad\quad\bm h\equiv\begin{bmatrix}
    1/2    & 0      & 0   & \cdots \\
    0      & 3/2    & 0   & \cdots \\
    0      & 0      & 5/2 & \\
    \vdots & \vdots &     & \ddots
    \end{bmatrix}\label{e3}
,\end{align}
and so we have the same positive-energy free spectrum for either $\sigma=\pm1$. The quantization scheme adopted for the $\sigma=-1$ case was chosen for this reason.

We move on to interacting ghost theories with only even powers of $\tilde{\bm{\phi}}$,
\begin{align}
\tilde{\textbf{H}}&=m\bm h+\sigma^\frac{k}{2}\frac{\lambda}{(2m)^\frac{k}{2}k!}\tilde{\bm{\phi}}^k,\quad\quad k=4,6,8,...
\end{align}
In this case it turns that  $\tilde{\textbf{H}}^\dagger=\tilde{\textbf{H}}$ and so these theories automatically have real spectra. These theories also have conserved ghost parity
\begin{align}
    [\bm\eta,\tilde{\textbf{H}}]=0
,\end{align}
and so produce Born rule probabilities from (\ref{e34}) that are automatically positive. The sign $\sigma^\frac{k}{2}$ assumes that $\lambda>0$ and it ensures that the spectrum is bounded from below.\footnote{This sign agrees with the use of the transformation $\phi\to i\phi$ and $\pi\to i\pi$  to relate the $\sigma=1$ and $\sigma=-1$ theories.}

The two matrices $\tilde{\textbf{H}}(\sigma=-1)$ and $\tilde{\textbf{H}}(\sigma=1)$ are not the same; they each have a banded structure, but every other band away from the diagonal differs by a sign. Nevertheless we find that $\tilde{\textbf{H}}(\sigma=-1)$ and $\tilde{\textbf{H}}(\sigma=1)$ are isospectral, by using the methods described below. These methods also show that the full propagators of the two theories only differ by a sign. General amplitudes must involve an even number of ghosts, $n_g$, and if $n_g/2$ is odd then the respective amplitudes in the two theories will differ in sign. There should be no physical consequences of this difference. We conclude that a ghost theory with ghost parity is physically equivalent to the corresponding normal theory. This includes ghost theories with ghosts interacting with normal particles, as long as the theory has $\tilde{\textbf{H}}^\dagger=\tilde{\textbf{H}}$ and $[\bm\eta,\tilde{\textbf{H}}]=0$.

Our focus shall be on the $k$-odd ghost theories that do not have these properties,
\begin{align}
\tilde{\textbf{H}}&=m\bm h+\frac{\lambda}{(2m)^\frac{k}{2}k!}\tilde{\bm{\phi}}^k\quad\quad k=3,5,7,....
\label{e9}\end{align}
Here either choice of the sign of $\lambda$ gives the same theory. Now the $\sigma=1$ and $\sigma=-1$ theories are distinctly different. For $\sigma=1$, it is well known that the odd power in the interaction gives a spectrum not bounded from below. For $\sigma=-1$ we finally arrive at a theory with $\tilde{\textbf{H}}^\dagger\neq\tilde{\textbf{H}}$. Here we might expect to add the problem of complex energies to the problem of a spectrum not bounded from below. But as we find in the next section, these expectations are deceiving.

\section{Numerical spectra and full propagators}\label{s4}
In this section we solve the eigenvalue problem numerically to determine the energy eigenvalues and eigenvectors. The $k$-odd ghost theory turns out to enjoy a positive real spectrum at both weak and strong coupling. We then go on to determine the full propagator, a correlation function in the 0+1d QFT. As we shall see in the next section, the energy eigenvalues and eigenvectors are also a prerequisite for defining a sensible inner product.

In a numerical study we truncate the Hilbert space by truncating the number of occupation number basis states $|n\rangle$ to a maximum $n_{\rm max}$, thus yielding a matrix $\tilde{\textbf{H}}$ of finite size. Then the behavior of the energy eigenvalues is studied as $n_{\rm max}$ is increased. For the $k$-odd ghost theory we find that for any finite size there is a set of eigenvalues that are the smallest in absolute value and that are also real and positive. This may comprise all of the eigenvalues if the coupling is sufficiently weak. For larger coupling some eigenvalues of higher absolute value come in complex-conjugate pairs, and these may be interspersed with the positive real eigenvalues, with the latter ceasing to exist at the highest absolute values. The number of positive eigenvalues increases as the size of the matrix increases. The value of any given positive eigenvalue quickly converges for increasing matrix size, while the value of any given complex eigenvalue is not stable and it eventually becomes real for increasing matrix size. Thus by increasing the matrix size, the number in the set of converged and positive eigenvalues can be made arbitrarily large. In this way we can argue that in the infinite size limit we converge to a purely real and positive spectrum, and that this is the spectrum of the theory.

We focus on the $k=3$ ghost theory in (\ref{e9}). To produce some large number of positive energy eigenvalues we find that not only the matrix size must be sufficiently high, but the numerical precision for the diagonalization algorithm must also be high. We use the Arnoldi method with 32 digits of precision and consider a matrix of size $600\times600$, which has a sparseness of 0.008. This produces positive values of $E_n-E_0$ up to $n=40$ for both weak and strong couplings. Table \ref{t1} shows results up to $n=9$ for both $\lambda=0.1$ and $\lambda=10$. We see that strong coupling tends to increase the spacing between the energy eigenvalues $E_n-E_0$. This provides another way to see why complex conjugate pairs do not occur, since that would require a merging of eigenvalues as the coupling is varied.
\begin{table}[h]
    \centering
    \begin{tabular}{|c|c|c|c|c|} \hline 
 & \multicolumn{2}{|c|}{$\lambda=1/10$}& \multicolumn{2}{|c|}{$\lambda=10$}\\ \hline 
 $n$& $E_n-E_0$& $Z_n$& $E_n-E_0$&$Z_n$\\ \hline 
         0&  0.0&  0.0006212113&  0.0& 0.1390183399\\ \hline 
         1&  1.0020710213&  $-$1.000547848&  2.4051860435& $-$1.0512830324\\ \hline 
         2&  2.0061953609&  0.0005480361&  5.208574457& 0.0522332501\\ \hline 
         3&  3.0123535145&  $-$1.881e-7&  8.2521064593& $-$0.0009618532\\ \hline 
         4&  4.0205264034&  1.0e-10&  11.4766245676& 1.17488e-5\\ \hline 
         5&  5.030695361&  &  14.848397294&  $-$1.143e-7\\ \hline 
         6&  6.0428421179&  &  18.3452838701& 1.0e-9\\ \hline 
         7&  7.0569487895&  &  21.951494836& \\ \hline 
         8&  8.072997863&  &  25.655134468& \\ \hline 
         9&  9.0909721854&  &  29.4468816584& \\ \hline
    \end{tabular}
    \caption{Model in (\ref{e9}) with $\sigma=-1$, $k=3$ and $m=1$. $Z_0$ is the first term in (\ref{e24}).}
    \label{t1}
\end{table}

The $Z_n$ show how the various states contribute to the full propagator. The full propagator can be expressed in terms of matrix elements involving the exact energy eigenstates (see \cite{Boo} for a discussion for 0+1d QFT),
\begin{equation}
\frac{\langle\bar0|T[\phi(t_b)\phi(t_a)]|\bar0\rangle}{\langle\bar0|\bar0\rangle}=\frac{\langle\bar 0|\phi|\bar0\rangle^2}{\langle\bar 0|\bar 0\rangle^2}+\sum_{n=1}^\infty Z_n D_F(t_b-t_a,E_n-E_0)
.\label{e24}\end{equation}
$D_F$ is the free 0+1d Feynman propagator,
\begin{align}
    D_F(\tau,\mu)&=\frac{1}{2\mu}e^{-i(\mu-i\epsilon)|\tau|}\nonumber\\
Z_n&=2(E_n-E_0)\frac{\langle\bar 0|\phi|\bar n\rangle\langle\bar n|\phi|\bar0\rangle}{\langle\bar0|\bar0\rangle\langle\bar n|\bar n\rangle}
.\end{align}
The eigenvalues and the eigenvectors determine the $Z_n$.  The first excited state produces the contribution $Z_1 D_F(\tau,E_1-E_0)$ where $Z_1$ is negative. At zero coupling this state is responsible for the free propagator, and this contribution still manages to dominate even at strong coupling. The $Z_n$ alternate in sign and the Table shows that the propagator can be determined accurately with a relatively small number of terms in the sum even at strong coupling.

The $k=5$ ghost theory gives qualitatively similar results. In Section \ref{s10} we shall consider a model with a normal field interacting with a ghost, and we also introduce models with more than one spatial point.

\section{New inner product}\label{s5}
We now address the question of whether the ghost theory supports another inner product that can be used to define a sensible Born rule. A new inner product, the $G$ inner product, can be defined as
\begin{align}
    \langle \psi|\chi\rangle_G= \bm\psi^\dagger \textbf{G} \bm\chi= (\textbf{G}\bm\psi)^\dagger  \bm\chi
.\end{align}
We are already insisting that $\textbf{G}^\dagger=\textbf{G}$. It is also useful to introduce $\tilde{\textbf{G}}\equiv\bm\eta\textbf{G}$, since $\tilde{\textbf{G}}$ is the additional factor that is inserted into the $\eta$ inner product to obtain the $G$ inner product, $\langle \psi|\chi\rangle_G= (\bm\eta\tilde{\textbf{G}}\bm\psi)^\dagger  \bm\chi$. The $G$ inner product must be preserved in time to respect unitary evolution. Since the $\eta$ inner product is already preserved in time we must have
\begin{align}
    [\tilde{\textbf{G}},\tilde{\textbf{H}}]=0
.\end{align}
Thus the energy eigenstates are also eigenstates of $\tilde{\textbf{G}}$. We can require that the corresponding eigenvalues of $\tilde{\textbf{G}}$ to be $\pm1$, and thus $\tilde{\textbf{G}}^2=1$. We can also insist that 
\begin{align}
    \tilde{\textbf{G}}^\dagger\textbf{G}&=\textbf{G}\tilde{\textbf{G}},\label{e25}\\
    \tilde{\textbf{H}}^\dagger\textbf{G}&=\textbf{G}\tilde{\textbf{H}},\label{e38}
\end{align}
so that both the $\tilde{\textbf{G}}$ and $\tilde{\textbf{H}}$ are self-adjoint with respect to the $G$ inner product. (\ref{e38}) is another statement that the $G$ inner product is preserved in time. $\tilde{\textbf{G}}$, like $\tilde{\textbf{H}}$, is also self-adjoint with respect to the $\eta$ inner product.

We find that the norms of the energy eigenstates alternate in sign just as they do in the occupation number basis,
\begin{align}
    \langle\bar m|\bar n\rangle=(\bm\eta\bm\psi^{(m)})^\dagger\bm\psi^{(n)}\propto(-1)^n\delta_{mn}
.\end{align} 
Thus the $\tilde{\textbf{G}}$ that will produce positive $G$-norms has $\tilde{\textbf{G}}\bm\psi^{(n)}=(-1)^n\bm\psi^{(n)}$. In other words $\tilde{\textbf{G}}$ is the ghost parity of the energy eigenstates (while $\bm\eta$ is the ghost parity of the occupation number basis). Such a $\tilde{\textbf{G}}$ can be realized by a matrix in the occupation number basis as
\begin{align}
\tilde{\textbf{G}}=\sum_{n}\frac{\bm\psi^{(n)}(\bm\eta\bm\psi^{(n)})^\dagger}{|(\bm\eta\bm\psi^{(n)})^\dagger\bm\psi^{(n)})|}
.\label{e36}\end{align}
The numerator of each term is a matrix, a product of a column vector with a row vector.

This $\tilde{\textbf{G}}$ satisfies the relations in (\ref{e25}-\ref{e38}) and as well we see that
\begin{align}
    \tilde{\textbf{G}}^2=\sum_{n}\frac{\bm\psi^{(n)}(\bm\eta\bm\psi^{(n)})^\dagger}{(\bm\eta\bm\psi^{(n)})^\dagger\bm\psi^{(n)}}=1
\end{align}
due to the completeness relation. It can also be seen that $[\tilde{\textbf{G}},\tilde{\textbf{H}}]=0$ since $\tilde{\textbf{H}}$ is also a combination of the $\bm\psi^{(n)}(\bm\eta\bm\psi^{(n)})^\dagger$. Most importantly it can be seen that $\textbf{G}=\bm\eta\tilde{\textbf{G}}$ is a positive-definite Hermitian matrix, and so all eigenvalues of $\textbf{G}$ are positive and all $G$-norms are positive.

Our discussion indicates that the $G$ inner product is the unique time-independent inner product with positive norms. The Born rule defined in terms of the $G$ inner product is
\begin{align}
    \textrm{Pr}(i\to f)=\frac{|(\textbf{G}\bm\psi_f)^\dagger\bm\psi_i|^2}{((\textbf{G}\bm\psi_i)^\dagger\bm\psi_i)((\textbf{G}\bm\psi_f)^\dagger\bm\psi_f)}
.\label{e20}\end{align}
We now have $0\leq\textrm{Pr}(i\to f)\leq1$ and along with $\sum_f\textrm{Pr}(i\to f)=1$, a sensible probability interpretation follows.

The matrix $\textbf{G}$ is nontrivial but it produces a trivial inner product for appropriately normalized energy eigenvectors,
\begin{align}
    \langle\bar m|\bar n\rangle_G=(\textbf{G}\bm\psi^{(m)})^\dagger  \bm\psi^{(n)}=\delta_{mn}
.\end{align}
If we write a general state in terms of the energy eigenstates,
\begin{equation}
    |\psi\rangle = \sum_{n\geq 0} \tilde{\psi}_n |\bar n \rangle,
\label{e16}\end{equation}
and define new column vectors such as $(\tilde{\bm{\psi}})_n =\tilde{\psi}_n$ then the $G$ inner product becomes
\begin{align}
    \langle \psi|\chi\rangle_G= \tilde{\bm\psi}^\dagger \tilde{\bm\chi}
.\end{align}
We recover the standard Dirac inner product in this way, and so the Born rule can now be written in the standard form
\begin{align}
    \textrm{Pr}(i\to f)=\frac{|\tilde{\bm\psi}_f^\dagger\tilde{\bm\psi}_i|^2}{(\tilde{\bm\psi}_i^\dagger\tilde{\bm\psi}_i)(\tilde{\bm\psi}_f^\dagger\tilde{\bm\psi}_f)}
.\end{align}
Of course this construction requires knowledge of the energy eigenstates, just as the matrix $\textbf{G}$ does. If $\tilde{\bm\psi}_f$ a member of a set of vectors that is orthonormal and complete in the conventional sense, then summing over this set gives $\sum_f\textrm{Pr}(i\to f)=1$. In other words, summing the probabilities over any complete set of states that is unitarily related to the set of energy eigenstates gives unity. It can also be said that any such set is orthonormal and complete with respect to the $G$ inner product.

The rule for ghost quantum mechanics is thus to first determine the energy eigenstates of the full Hamiltonian, use this as a basis for which to express arbitrary states, and then calculate probabilities via the normal form of the Born rule. If a different basis is used then the form of the Born rule changes, as in (\ref{e20}) where states are expressed in terms of the occupation number basis.

\section{Wave functions}
In this section we develop a coordinate representation, and from that we can obtain the wave functions for the energy eigenstates. We first discuss how to obtain observables in a ghost theory in general. Given an operator $A$, the eigenvalue equation $A|\lambda\rangle=\lambda|\lambda\rangle$ when translated to matrix notation becomes $\tilde{\textbf{A}}\bm\psi^{\lambda}=\lambda\bm\psi^{\lambda}$. We are guaranteed to have real eigenvalues if  $\tilde{\textbf{A}}$ is Hermitian,  $\tilde{\textbf{A}}^\dagger=\tilde{\textbf{A}}$, and this means that the matrix $\textbf{A}$ and the operator $A$ are self-adjoint. Thus self-adjoint operators can be observables. The Hamiltonian is not such an example since the matrix $\textbf{H}$ is Hermitian while the matrix $\tilde{\textbf{H}}$ is self-adjoint. The fact that real eigenvalues emerged for $\tilde{\textbf{H}}$ in the infinite size limit is apparently an exception to the rule.

Let us find suitable operators $q$ and $p$ to serve as position and momentum operators. The matrix form of the commutation relation $[q,p]=i$ is
\begin{align}
    \textbf{q}\bm\eta\textbf{p}-\textbf{p}\bm\eta\textbf{q}&=i\bm\eta,\\
\textrm{or}\quad    \tilde{\textbf{q}}\tilde{\textbf{p}}-\tilde{\textbf{p}}\tilde{\textbf{q}}&=i
.\end{align}
There are two solutions to the latter relation, ($\tilde{\textbf{q}}=\bm\phi$, $\tilde{\textbf{p}}=\sigma\bm\pi$) and ($\tilde{\textbf{q}}=i\tilde{\bm\phi}$, $\tilde{\textbf{p}}=\sigma i\tilde{\bm\pi}$). Both sets of solutions to the commutation equations are such that $\tilde{\textbf{q}}$ and $\tilde{\textbf{p}}$ are Hermitian, thus producing real eigenvalues to the matrix eigenvalue equations. Correspondingly the self-adjoint operators $q$ and $p$ for the two choices have matrices ($\textbf{q}=\tilde{\bm\phi}$, $\textbf{p}=\sigma\tilde{\bm\pi}$) or ($\textbf{q}=i\bm\phi$, $\textbf{p}=\sigma i\bm\pi$). The latter corresponds to the Dirac-Pauli choice \cite{Salvio:2015gsi,Strumia:2017dvt,Salvio:2018crh}.\footnote{Unlike those references we do not use derivative representations of operators. The matrices $\bm\phi$, $\bm\pi$, $\tilde{\bm\phi}$, $\tilde{\bm\pi}$ are shown in (\ref{e28}) and (\ref{e32}).} (Two solutions are made possible by the presence of $\bm\eta$ for $\sigma=-1$; for a normal $\sigma=1$ theory there is only one solution $\textbf{q}=\bm\phi$, $\textbf{p}=\bm\pi$.)

We will first consider the Dirac-Pauli choice and thus the following eigenvalue equation, which determines a set of real positions $x$,
\begin{align}
    (i\tilde{\bm\phi})\bm\psi^{(x)}=x\bm\psi^{(x)}
.\label{e14}\end{align}
The eigenvector $\bm\psi^{(x)}$ is representing the position eigenstate $|x\rangle$, and with these eigenvectors we can find the inner product in the coordinate representation $\langle x'|x\rangle$. This inner product is not diagonal, and the overall signs of the eigenvectors in (\ref{e14}) can be chosen such that
\begin{align}
    \langle x'|x\rangle=(\bm\eta\bm\psi^{(x')})^\dagger\bm\psi^{(x)}=\delta_{-x,x'}
\label{e19}.\end{align}
This agrees with \cite{Salvio:2015gsi,Strumia:2017dvt,Salvio:2018crh}. The implication is that
\begin{align}
    \bm\eta\bm\psi^{(x)}=\bm\psi^{(-x)}
\end{align}
and thus $\bm\eta$ acts as a normal parity operator on these position eigenvectors.

The position eigenvectors allow us to produce a wave function (a function of $x$) for any of the energy eigenstates $|\bar n\rangle$ as
\begin{align}
    \langle x|\bar n\rangle=(\bm\eta\bm\psi^{(x)})^\dagger\bm\psi^{(n)}
.\end{align}
The corresponding momentum space wave functions are determined similarly starting with
\begin{align}
    (\sigma i\tilde{\bm\pi})\bm\psi^{(p)}=p\bm\psi^{(p)}
.\end{align}

Truncating $i\tilde{\bm\phi}$ to a finite size in the eigenvalue equation (\ref{e14}) yields a finite set of values for the position, with these values occurring symmetrically around $x=0$. Taking the matrix $i\tilde{\bm\phi}$ arbitrarily large will populate an arbitrarily large range of $x$, arbitrarily densely with points. The result is that the discrete set of values for $|\langle x|\bar n\rangle|$ converges to a smooth function over the real line. We find that $|\langle x|\bar n\rangle|$ falls as a Gaussian for large $|x|$, or even faster for strong coupling, and that its values become negligible before reaching the maximum $|x|$ that exists for a finite size $i\tilde{\bm\phi}$.

\begin{figure}
    \centering
    \includegraphics[width=0.6\linewidth]{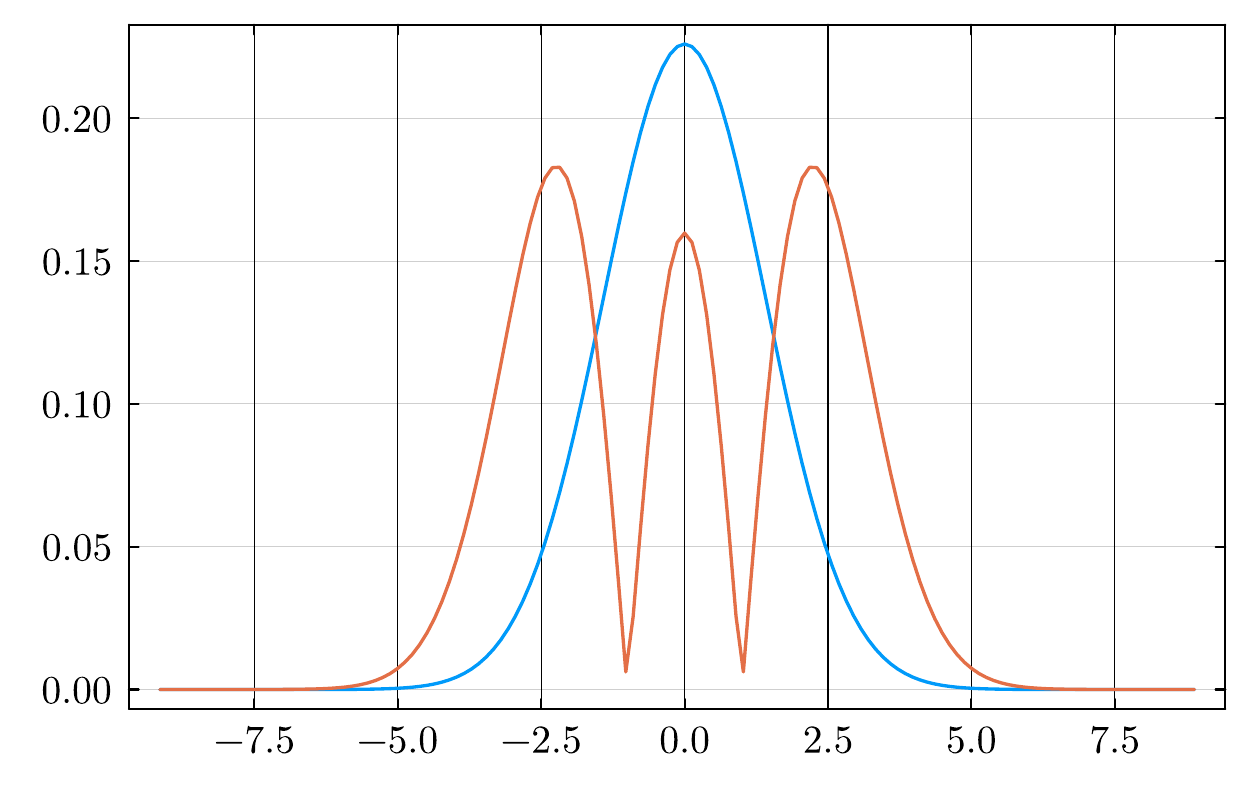}
    \caption{$|\langle x|\bar 0\rangle|=|\langle p|\bar 0\rangle|$ (blue) and $|\langle x|\bar 2\rangle|=|\langle p|\bar 2\rangle|$ (orange) for $\lambda=0$.}
    \label{f1}
\end{figure}

\begin{figure}
    \centering
    \includegraphics[width=0.6\linewidth]{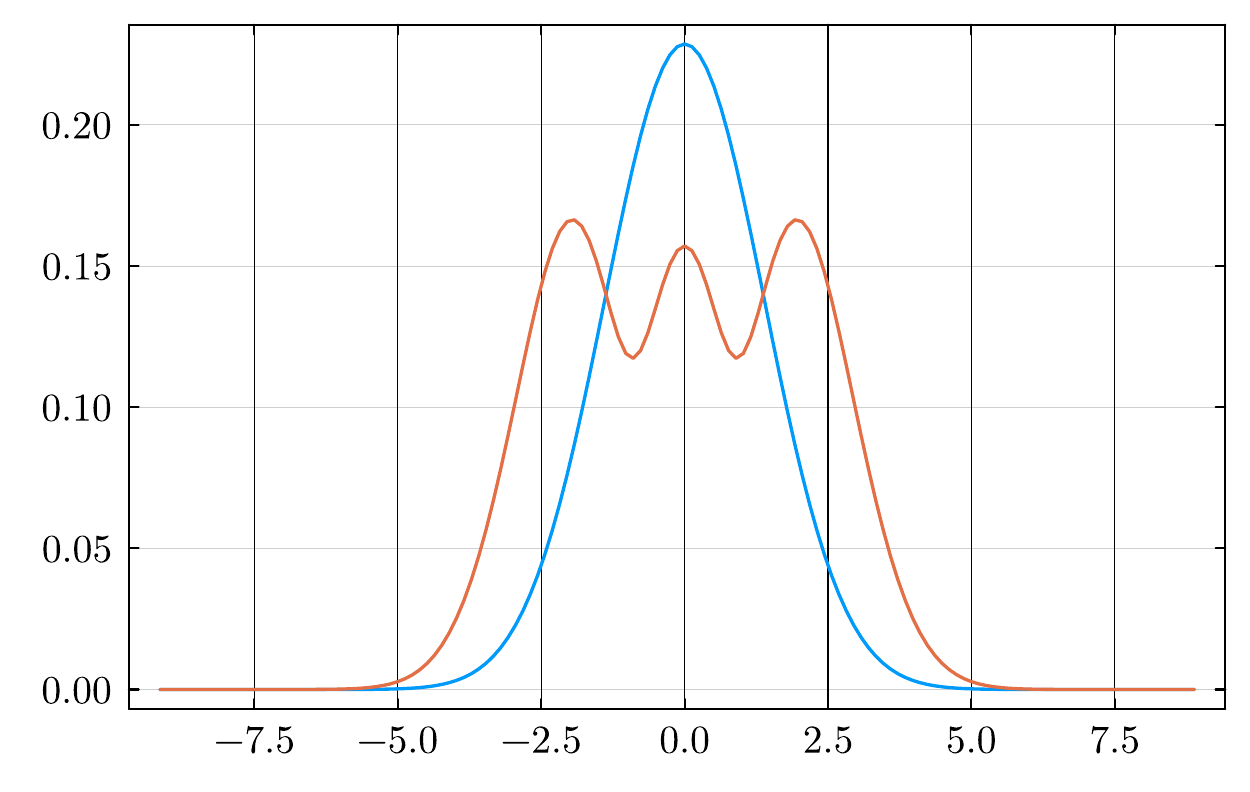}
    \caption{$|\langle x|\bar 0\rangle|$ (blue) and $|\langle x|\bar 2\rangle|$ (orange) for $\lambda=1$.}
    \label{f2}
\end{figure}

\begin{figure}
    \centering
    \includegraphics[width=0.6\linewidth]{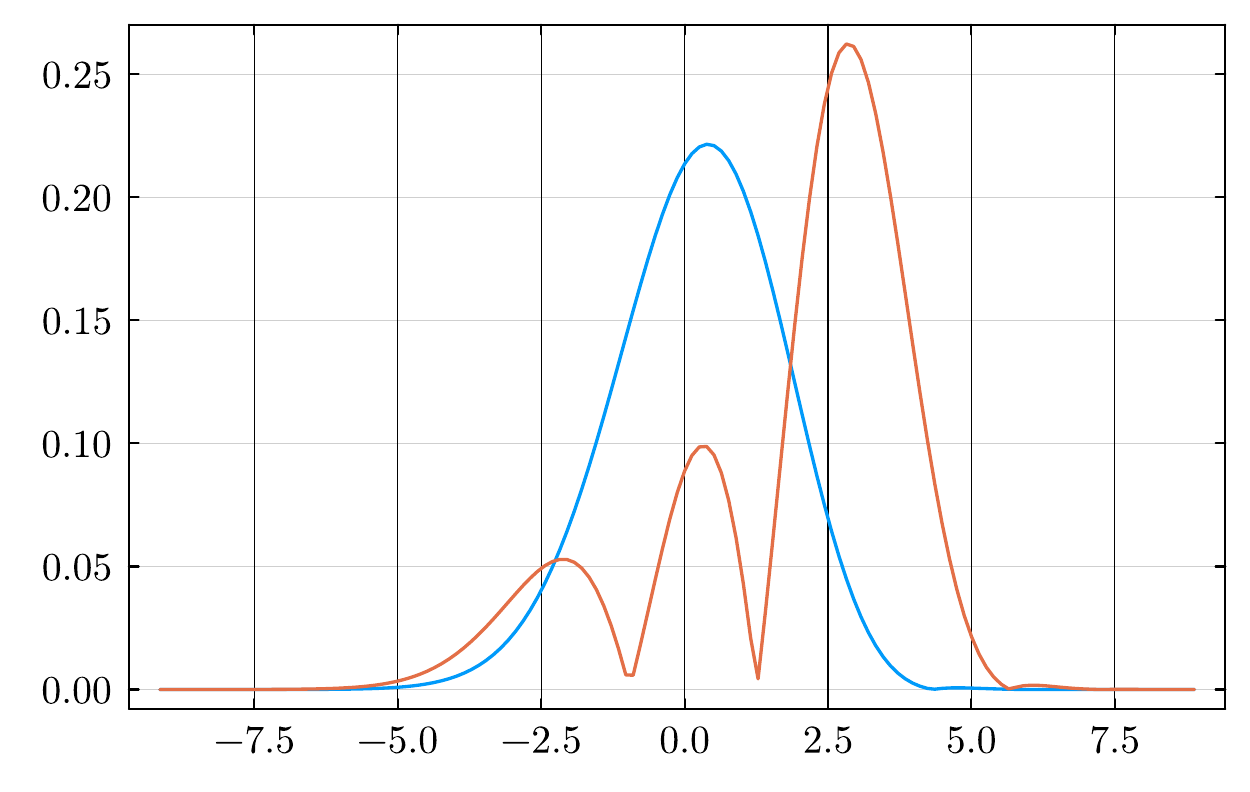}
    \caption{$|\langle p|\bar 0\rangle|$ (blue) and $|\langle p|\bar 2\rangle|$ (orange) for $\lambda=1$.}
    \label{f3}
\end{figure}

Since we are forming wave functions for the exact energy eigenstates, these wave functions incorporate the effects of interactions. We consider the $k=3$ ghost theory in (\ref{e9}). At zero coupling the wave functions $|\langle x|\bar n\rangle|$ and $|\langle p|\bar n\rangle|$ are the same, as shown in Fig.~\ref{f1}. At $\lambda=1$ we show $|\langle x|\bar n\rangle|$ in Fig.~\ref{f2}, where we see that interactions have caused the wave functions to become more localized in coordinate space. We would then expect the momentum space wave functions to be more spread out. In Fig.~\ref{f3} we see this, but also another phenomenon occurs. The momentum space wave functions are showing the explicit breaking of normal parity due to interactions. The latter occurs since we have both $\bm\eta\bm\psi^{(p)}=\bm\psi^{(-p)}$ and $[\bm\eta,\tilde{\textbf{H}}]\neq0$. If we had instead chosen the other solution, that is $\textbf{q}=\tilde{\bm\phi}$ and $\textbf{p}=\sigma\tilde{\bm\pi}$, then the parity violation shows up in the coordinate-space wave functions instead. Again this only occurs for non-vanishing coupling.



We note that the matrix $\langle x'|x\rangle=\delta_{-x,x'}$ in (\ref{e19}) has half of its eigenvalues negative, just as for $\langle n'|n\rangle$. The effect of the $G$ inner product is to cause $\langle x'|x\rangle_G$ to have only positive eigenvalues, while maintaining $|\langle x|\bar n\rangle_G|=|\langle x|\bar n\rangle|$. The $G$ inner product brings in the effects of interactions and so $\langle x'|x\rangle_G$ is a nontrivial matrix. That is the $|x\rangle$ do not form an orthonormal set with respect to the $G$ inner product. It is only in the case of vanishing coupling that the standard result $\langle x'|x\rangle_G=\delta_{x,x'}$ is recovered.

We may also consider the time evolution of wave functions. The time evolution of the arbitrary state in (\ref{e16}) is
\begin{align}
    |\psi(t)\rangle=\sum_n \tilde{\psi}_n e^{-i(E_n-E_0)t} |\bar n\rangle
.\label{e35}\end{align}
An evolving wave function is obtained by taking an inner product with a non-time-evolving position eigenstate,
\begin{align}
    \langle x|\psi(t)\rangle=\sum_n  \tilde{\psi}_n e^{-i(E_n-E_0)t} \langle x|\bar n\rangle
.\label{e27}\end{align}
We can obtain the evolving probability density $P_\psi (x,t)$ from the Born rule in (\ref{e20}),
\begin{align}
    P_\psi (x,t)=\frac{|\langle x|\psi(t)\rangle_G|^2}{\langle x|x\rangle_G\langle\psi|\psi\rangle_G}
,\label{e2}\end{align}
since the position eigenvectors are expressed in the occupation number basis. The time dependence is all in the numerator, and since $\textbf{G}=\bm\eta\tilde{\textbf{G}}$, the $\tilde{\textbf{G}}$ introduces a $(-1)^n$ factor in the $G$-norm version of (\ref{e27}),
\begin{align}
    \langle x|\psi(t)\rangle_G=\sum_n \tilde{\psi}_n (-1)^n e^{-i(E_n-E_0)t} \langle x|\bar n\rangle
.\label{e26}\end{align}

An evolving probability density in standard quantum mechanics is $|\psi(x,t)|^2$, where $\psi(x,t)$ is typically calculated with the Schrodinger equation. For the ghost theory we have not obtained the analog of the Schrodinger equation. And our result for $P_\psi (x,t)$ differs from standard quantum mechanics in other ways. One is the appearance of the $(-1)^n$ factor in (\ref{e26}). Another is the fact that the position eigenstates are neither orthonormal nor complete with respect to the $G$ inner product, for non-vanishing coupling. Thus if we sum $P_\psi (x,t)$ in (\ref{e2}) over all values of $x$, this sum need not be unity or time independent.

Finally we note that for the finite dimensional truncation that we are using in this section, complex energies will appear in the spectrum for large enough coupling. In Section \ref{s8} we shall obtain the $G$ inner product for this case.
 
\section{The PT connection}

The construction of a sensible inner product from the eigenstates of a non-Hermitian Hamiltonian is central to the study of PT-symmetric Hamiltonians \cite{Bender:2023cem}. The reader may have noticed the similarities in our discussion of the inner product, where our operator $\tilde{G}$ corresponds to the operator $C$ in those studies, e.g.~\cite{BM1,BM2,Mannheim:2009zj}. But PT-symmetric quantum mechanics is typically approached as a complex extension of ordinary quantum mechanics via the construction of a complex coordinate-space representation. We are instead starting with a Hermitian Hamiltonian and a canonical quantization, where it is instead the effect of negative norms that leads to the consideration of a non-Hermitian (but self-adjoint) matrix $\tilde{\textbf{H}}$.

Canonical quantization can also be applied to intrinsically non-Hermitian, PT-symmetric theories. In particular a PT-symmetric Hamiltonian is obtained from our $k$-odd Hamiltonian in (\ref{e9}) with the transformation $\phi\to i\phi$ and $\pi\to i\pi$. The kinetic terms are then positive and a $\sigma=1$ canonical quantization can proceed, resulting in the norm $\langle n' | n \rangle = \delta_{nn'}$ for the occupation number basis. But since this $k$-odd Hamiltonian with imaginary coupling is explicitly non-Hermitian, this norm is not preserved under time evolution. Nevertheless the Hamiltonian can be diagonalized and the energy eigenvalues and eigenvectors can be determined. The result is that this PT Hamiltonian produces a spectrum that is isospectral to the ghost Hamiltonian (\ref{e9}).

To make sense of this PT theory, an inner product must be found to be consistent with unitary evolution. It is simplest to require that the bra-state $\langle\psi|$ of all matrix elements be represented by $(\bm\eta\bm\psi)^\dagger$ rather than $\bm\psi^\dagger$. Here we are introducing $\bm\eta\equiv\bm\eta_{-}$ even though we are working in a $\sigma=1$ theory. Similarly the right and left energy-eigenstates of the non-Hermitian matrix Hamiltonian $\textbf{H}$ (no tilde) can be chosen to be
\begin{equation}
    \textbf{H} \bm{\psi}^{(n)} =E_n \bm{\psi}^{(n)}\quad\quad   (\bm\eta\bm\psi^{(n)})^\dagger \textbf{H}  =E_n^* (\bm\eta\bm\psi^{(n)})^\dagger
.\end{equation}
The inner product, the same as in (\ref{e37}), is now conserved under time evolution since we find that $\textbf{H}^\dagger\bm\eta=\bm\eta\textbf{H}$. In other words, $\textbf{H}$ is self-adjoint with respect to the inner product. All the consequences of unitary evolution follow. This setup for the PT theory is similar to the starting point of the ghost theory. But in the ghost theory the unitary evolution was built-in and uniquely defined by the theory, while in the PT theory the unitary evolution requires some appropriate choice of the bra-states.

At this stage both theories have negative norms. To obtain an inner product that gives a sensible Born rule, the $G$ inner product defined by $\textbf{G}=\bm\eta\tilde{\textbf{G}}$, with $\tilde{\textbf{G}}$ in (\ref{e36}), also works for the corresponding PT theory. This type of construction is known from PT studies, as is the fact that the exact energy eigenstates are required to obtain the desired inner product \cite{Bender:2023cem}. But for the PT theory there appears to be an ambiguity; should the $G$ inner product be used for all matrix elements, or only for the Born rule? To pursue a correspondence with the ghost theory, we use the $G$ inner product only for the Born rule.

If we proceed to the calculation of the propagator in the PT theory, the results can again be expressed in terms of the single field matrix elements $\langle n'|\phi|n\rangle$. But the matrix representation of these matrix elements has changed from $\bm\psi^{(n')^\dagger}\bm\phi\bm\psi^{(n)}$ in the ghost theory to $(\bm\eta\bm\psi^{(n')})^\dagger\bm\phi\bm\psi^{(n)}$ in the PT theory. The matrix $\bm\phi$ has also changed, now being given in (\ref{e28}) with $\sigma=1$ rather than with $\sigma=-1$. We find that the net result of these changes is for the $Z_n$ factors in the full propagator to all change sign relative to the ghost theory. This is the only change and so the full propagators of the two theories only differ by a sign.

Similarly amplitudes with $n_g$ fields will only differ by a factor of $i^{n_g}$ in the two theories, due to the simple $\phi\to i\phi$ and $\pi\to i\pi$ transformation relating the two theories. These factors should not have any physical consequence. Thus we are being led to conclude that the $k=3$ ghost theory and a certain non-Hermitian PT theory give equivalent quantum field theories.

It is instructive to explore further which PT theories are equivalent to ghost theories. Numerical results for spectra are readily available for the following massless PT theories where a choice of coupling has been made, 
\begin{align}
    H=\pi^2+\phi^n(i\phi)^\varepsilon\quad\quad n=2,4,6,\dots
\end{align}
These theories are deformations of normal theories by continuously changing the parameter $0<\varepsilon<2$. The positive spectra for the $n=2$ and $4$ theories are shown in \cite{Bender:2023cem}. When $\varepsilon=1$ and for any of the $n$ listed we can use our methods of canonical quantization (with $\sigma=1$) and matrix methods to determine the spectra of these PT theories. The spectra determined in this way match those shown in \cite{Bender:2023cem}. Similarly we can determine the spectra of the corresponding ghost theories,
\begin{align}
    H=-\pi^2+\phi^k\quad\quad k=3,5,7,\dots
\end{align}
Now the quantization proceeds with $\sigma=-1$ and the spectra again match.

Another set of PT theories is given by
\begin{align}
    H=\pi^m+\phi^2(i\phi)^\varepsilon\quad\quad m=4,6,8,\dots
\end{align}
The spectra of these theories with $0<\varepsilon<2$ and for $m=2$ and $4$ are shown in \cite{Bender:2023cem}. The related ghost theories with $\varepsilon=1$ are
\begin{align}
    H=(-1)^\frac{m}{2}\pi^m+\phi^3\quad\quad m=4,6,8,\dots
\end{align}
By quantizing the $\varepsilon=1$ PT theories with $\sigma=1$ and the ghost theories with $\sigma=-1$, we again find that all the spectra match.

We have found matching spectra for the choice $\varepsilon=1$ in the PT theory. In the terminology of the PT literature, these particular theories have Stokes wedges that continue to include the real axis. It is only this particular type of PT theory that is connected to ghost theories. In our canonical approach to PT theories, matrix eigenvalue equations can again be used to determine the coordinate and momentum representations. The commutation relation $[q,p]=i$ in matrix form for $\sigma=1$ is simply ${\textbf{q}}{\textbf{p}}-{\textbf{p}}{\textbf{q}}=i$ and this again has two solutions ($\textbf{q}={\bm\phi}$, $\textbf{p}={\bm\pi}$) or ($\textbf{q}=i\bm\eta\bm\phi$, $\textbf{p}=i\bm\eta\bm\pi$). The second choice is made possible because of the appearance of $\bm\eta$ in the description of the PT theory. Real values of $x$ and $p$ and normalizable wave-functions are obtained with no need of a complex extension. The effect of interactions in a $k$-odd PT theory again violates normal parity, in the momentum wave functions for the first choice and the coordinate wave functions for the second choice.

\section{Complex-conjugate pairs of states}\label{s8}
In this section and the next we consider the case where the energy spectrum includes complex-conjugate pairs. We have noted that this occurs for the $k$-odd ghost theory when the Hilbert space is made finite dimensional, by truncating infinite matrices to a finite size. It also requires that the coupling be sufficiently large. Complex-conjugate pairs of states are of special interest in a ghost QFT in 3+1d. There they show up in the one-loop corrected ghost propagator as complex-conjugate poles. The particles in the loop are normal, and there is an energy degeneracy between the ghost and a two particle state. In the truncated 0+1d theory, the appearance of the complex-conjugate states is due to the merging of a ghost energy level and a normal energy level as the coupling is increased. This model provides us with the opportunity to study such states in a simple context.

We can label a particular pair of states with complex-conjugate energy eigenvalues by $p$ and $p+1$, and the whole set of such pairs by $\{p\}$ with $p=1,3,5,\dots$. The diagonal norms $(\bm\eta\bm\psi^{(p)})^\dagger \bm\psi^{p}$ and $(\bm\eta\bm\psi^{(p+1)})^\dagger \bm\psi^{p+1}$ are vanishing as we have already explained below (\ref{e4}). Instead there are off-diagonal inner products,
\begin{align}
\langle\overline p|\overline{p+1}\rangle&=(\bm\eta\bm\psi^{(p)})^\dagger\bm\psi^{p+1}=e^{i\theta_p},\nonumber\\\langle\overline {p+1}|\overline{p}\rangle&=(\bm\eta\bm\psi^{(p+1)})^\dagger \bm\psi^{p}=e^{-i\theta_p}.
\end{align}
We find that coordinate-space wave functions for these states, $|\langle x|\bar p\rangle|$ and $|\langle x|\overline{p+1}\rangle|$, are not symmetrical about $x=0$, and are parity reflections of each other. The set of all states is now $\{n\}=\{q\}\cup\{p\}$. $\{q\}$ with $q=0,1,2,\dots$ labels the set of states with real energies, ordered by energy. These states have alternating sign norms as before,
\begin{align}
    \langle\bar q|\bar q\rangle=(\bm\eta\bm\psi^{(q)})^\dagger\bm\psi^{q}=(-1)^q
.\end{align}

It is convenient to consider combinations of energy eigenstates so that we can replace the $\{p\}$ states by states with diagonal norms. We define the following,
\begin{align}
  \bm\psi_\alpha^{(p)}= e^{i\alpha_p}\bm\psi^{(p+1)}+e^{-i\alpha_p}\bm\psi^{(p)},\\
  \bm\psi_\beta^{(p)}= e^{i\beta_p}\bm\psi^{(p+1)}+e^{-i\beta_p}\bm\psi^{(p)}.
\end{align}
By setting $\beta_p=\pi/2-\theta_p-\alpha_p$, the $\alpha$- and $\beta$-states are orthogonal, $\bm\psi_\alpha^{(p)\dagger}\bm\eta\bm\psi_\beta^{(p)}=0$, and have norms
\begin{align}
&\langle \bar\alpha_p|\bar\alpha_p\rangle = \bm\psi_\alpha^{(p)\dagger}\bm\eta\bm\psi_\alpha^{(p)}=2\cos(2\alpha_p+\theta_p),\\ &\langle \bar\beta_p|\bar\beta_p\rangle = \bm\psi_\beta^{(p)\dagger}\bm\eta\bm\psi_\beta^{(p)}=-2\cos(2\alpha_p+\theta_p).
\end{align}
Thus one and only one of these states has a negative norm. Since the inner product in general is preserved under time evolution, so also are these norms.

To complete the definition of the $\alpha$- and $\beta$-states we choose $\alpha_p$ such that the state $|\bar \alpha_p\rangle$ has no overlap with $\phi|\bar 0\rangle$, that is\footnote{As a function of $\alpha_p$, $\bm\psi_\alpha^{(p)\dagger}\bm\phi\bm\psi^{(0)}$ turns out to be purely real with a zero in the range $0<\alpha_p<\pi$.}
\begin{align}
    \langle \bar \alpha_p|\phi|\bar 0\rangle=\bm\psi_\alpha^{(p)\dagger}\bm\phi\bm\psi^{(0)}=0
.\end{align}
The meaning of this condition is that the state $|\bar \alpha_p\rangle$ does not contribute to the spectral function in the spectral representation of the propagator. In a higher dimensional QFT another implication of this condition is that the state $|\bar \alpha_p\rangle$ is not an asymptotic state. Asymptotic states are those that can participate in scattering experiments, that is they are states that can appear in scattering matrix elements as extracted from correlation functions by applying the LSZ reduction formula. More simply, asymptotic states must be able to propagate to or from the interaction region, and for this they must contribute to the spectral function. Although asymptotic states have little meaning for 0+1d QFT, we shall use this terminology and say that the $\beta$-states are asymptotic states, while the $\alpha$-states are not.

We now wish to utilize any freedom in the definition of the $G$ inner product to produce positive $G$-norms. The original norms are negative for odd $q$ in $\{q\}$ and they are negative for either the $\alpha$-state or $\beta$-state for each $p$ in $\{p\}$. As before we require $[\tilde{\textbf{G}},\tilde{\textbf{H}}]=0$, but in the present basis $\tilde{\textbf{H}}$ is not diagonal in each $(\bar\alpha_p,\bar\beta_p)$ subspace. Thus for each $p$ in $\{p\}$ there is only one sign to choose, $\tilde{\textbf{G}}(\bm\psi_\alpha^{(p)},\bm\psi_\beta^{(p)})=\pm(\bm\psi_\alpha^{(p)},\bm\psi_\beta^{(p)})$. We may choose this to be the sign of $\langle \bar\beta_p|\bar\beta_p\rangle$, and as before we choose $\tilde{\textbf{G}}\bm\psi^{(q)}=(-1)^q\bm\psi^{(q)}$. With these choices all states have positive $G$-norm except for the $\alpha$-states. In other words, all the asymptotic states have positive $G$-norm.

The expression for $\tilde{\textbf{G}}$ is
\begin{align}
\tilde{\textbf{G}}=\sum_{\{q\}}\frac{\bm\psi^{(q)}(\bm\eta\bm\psi^{(q)})^\dagger}{|(\bm\eta\bm\psi^{(q)})^\dagger\bm\psi^{(q)})|}+\sum_{\{p\}}\left[-\frac{\bm\psi_\alpha^{(p)}(\bm\eta\bm\psi_\alpha^{(p)})^\dagger}{|(\bm\eta\bm\psi_\alpha^{(p)})^\dagger\bm\psi_\alpha^{(p)}|}+
\frac{\bm\psi_\beta^{(p)}(\bm\eta\bm\psi_\beta^{(p)})^\dagger}{|(\bm\eta\bm\psi_\beta^{(p)})^\dagger\bm\psi_\beta^{(p)}|}\right]
.\end{align}
This $\tilde{\textbf{G}}$ satisfies $\tilde{\textbf{G}}^2=1$ and the relations in (\ref{e25}-\ref{e38}). This $\tilde{\textbf{G}}$ works for any finite size truncation of our Hamiltonians. For sufficiently weak coupling $\{p\}$ becomes empty and we recover (\ref{e36}). For vanishing coupling $\tilde{\textbf{G}}\to\bm\eta$.

An arbitrary state in the space of asymptotic states is
\begin{equation}
    |\psi\rangle = \sum_{\{q\}} \tilde{\psi}_q |\bar q \rangle+\sum_{\{p\}} \tilde{\psi}_p |\bar \beta_p \rangle
.\end{equation}
We assume that the basis states are normalized to unit norm. If we collect the coefficients appearing here into one column vector $\tilde{\bm\psi}$, the $G$-norm in terms of such column vectors is $\langle \psi|\chi\rangle_G= \tilde{\bm\psi}^\dagger \tilde{\bm\chi}$ as before. Thus once again, when using these column vectors, the Born rule can be written in the standard form involving the Dirac inner product.

We remind the reader that our focus has been on models with infinite dimensional Hilbert spaces, and for these the spectra are purely real and positive. Then the set $\{p\}$ is empty to begin with. We may also consider how the truncated models approach to the infinite size limit. As the finite size of the truncated model is increased, the $\{p\}$ states will have ever increasing complex energy. For any finite energy, there is a sufficiently large size of truncation such that the space spanned by states below that energy will be a positive $G$-norm subspace.

\section{Full propagator again}
We would like to extend the result for the full propagator in (\ref{e24}) to include contributions from complex-conjugate pairs of states. We first convert the defining correlation function into matrix notation. The tilde-matrix representation of operators is convenient since $\tilde{\bm\phi}(t)=\bm\eta e^{i\tilde{\bm H}^\dagger t_a}\bm\phi e^{-i\tilde{\bm H} t_a}=e^{i\tilde{\bm H} t_a}\tilde{\bm\phi} e^{-i\tilde{\bm H} t_a}$. We write
\begin{align}   \langle\bar0|T[\phi(t_b)\phi(t_a)]|\bar0\rangle=\bm\psi^{(0)\dagger}T[\bm\phi(t_b)\bm\eta\bm\phi(t_a)]\bm\psi^{(0)}=(\bm\eta\bm\psi^{(0)})^\dagger T[\tilde{\bm\phi}(t_b)\tilde{\bm\phi}(t_a)]\bm\psi^{(0)}
.\end{align}
We next insert unity in the form of the completeness relation between the two operators. Here we give two versions of the completeness relation using the decomposition of states, $\{n\}=\{q\}\cup\{p\}$.
\begin{align}
&\sum_{\{q\}}\frac{\bm\psi^{(q)}(\bm\eta\bm\psi^{(q)})^\dagger}{(\bm\eta\bm\psi^{(q)})^\dagger\bm\psi^{(q)}}+\sum_{\{p\}}\left[\frac{\bm\psi^{(p+1)}(\bm\eta\bm\psi^{(p)})^\dagger}{(\bm\eta\bm\psi^{(p)})^\dagger\bm\psi^{(p+1)}}+
\frac{\bm\psi^{(p)}(\bm\eta\bm\psi^{(p+1)})^\dagger}{(\bm\eta\bm\psi^{(p+1)})^\dagger\bm\psi^{(p)}}\right]=1
\label{e18}\\
&\sum_{\{q\}}\frac{\bm\psi^{(q)}(\bm\eta\bm\psi^{(q)})^\dagger}{(\bm\eta\bm\psi^{(q)})^\dagger\bm\psi^{(q)}}+\sum_{\{p\}}\left[\frac{\bm\psi_\alpha^{(p)}(\bm\eta\bm\psi_\alpha^{(p)})^\dagger}{(\bm\eta\bm\psi_\alpha^{(p)})^\dagger\bm\psi_\alpha^{(p)}}+\frac{\bm\psi_\beta^{(p)}(\bm\eta\bm\psi_\beta^{(p)})^\dagger}{(\bm\eta\bm\psi_\beta^{(p)})^\dagger\bm\psi_\beta^{(p)}}\right]=1
\label{e31}\end{align}
Now consider the complex-conjugate pair of states $\bm\psi^{(p)}$ and $\bm\psi^{(p+1)}$ with energies $E_p^r-iE_p^i$ and $E_p^r+iE_p^i$ with $E_p^i>0$, and the contribution from the corresponding two terms in the completeness relation in (\ref{e18}).  For one of these two terms, with $t_b>t_a$, we have the product of the two matrix elements on the following two lines,
\begin{align}
(\bm\eta\bm\psi^{(p)})^\dagger e^{i\tilde{\bm H} t_a}\tilde{\bm\phi} e^{-i\tilde{\bm H} t_a}\bm\psi^{(0)}=e^{i(E_p^r+iE_p^i-E_0)t_a}\bm\psi^{(p)\dagger} \bm\phi\bm\psi^{(0)}\nonumber\\
(\bm\eta\bm\psi^{(0)})^\dagger e^{i\tilde{\bm H} t_b}\tilde{\bm\phi} e^{-i\tilde{\bm H} t_b}\bm\psi^{(p+1)}=e^{-i(E_p^r+iE_p^i-E_{0})t_b}\bm\psi^{(0)\dagger} \bm\phi\bm\psi^{(p+1)}\nonumber
\end{align}
Doing this for both terms and for both orderings of $t_b$ and $t_a$ gives
\begin{equation}
\left.\frac{\langle\bar0|T[\phi(t_b)\phi(t_a)]|\bar0\rangle}{\langle\bar0|\bar0\rangle}\right|_\textrm{c.c.pairs}=\sum_{\{p\}} \left(Z_p D_F(t_b-t_a,E_p-E_0)+Z_p^*D_F(t_b-t_a,E_p^*-E_0)\right)
\label{e22},\end{equation}
where $D_F$ is the Feynman propagator as before and
\begin{align}
Z_p&=2(E_p^r-E_0)\frac{\langle\bar 0|\phi|\overline{p+1}\rangle\langle\bar p|\phi|\bar0\rangle}{\langle\bar0|\bar0\rangle\langle\bar p|\overline{p+1}\rangle}
.\label{e30}\end{align}
These $Z_p$ are complex with phases originating in both the numerator and denominator of (\ref{e30}). As before, the magnitude of the $Z_p$'s tend to fall exponentially with the size of $E_p^r-E_0$.

Standard manipulations involving the field commutator $\langle\bar0|[\phi(t_b),\phi(t_a)]|\bar0\rangle$ can be used to obtain the sum rule \cite{Boo,Kubo:2024ysu},
\begin{align}
    \sum_{\{q\}} Z_q+\sum_{\{p\}} (Z_p+Z_p^*)=\sigma
.\end{align}
We also have a result that originates from the use of the completeness relation in (\ref{e31}),
\begin{align}
    Z_p+Z_p^*=2(E_p^r-E_0)\frac{\langle\bar 0|\phi|\bar\beta_p\rangle\langle\bar\beta_p|\phi|\bar0\rangle}{\langle\bar0|\bar0\rangle\langle\bar\beta_p|\bar\beta_p\rangle}
.\end{align}
The $\bar\alpha_p$ state does not contribute. $\langle\bar 0|\phi|\bar\beta_p\rangle=\langle\bar\beta_p|\phi|\bar0\rangle$ is real and the sign of $Z_p+Z_p^*$ is determined by the sign of $\langle\bar\beta_p|\bar\beta_p\rangle$.

The new propagator contributions in (\ref{e22}) contain factors like $e^{\pm E_p^i t}$. But these exponential factors do not show up in inner products since the latter are time independent. This unusual behavior in the temporal version of the full propagators is consistent with unitarity and a sensible Born rule. Nevertheless it suggests that when complex conjugate pairs of states exist, as in the truncated models, then perturbation theory should be strictly applied, meaning that perturbation theory is to make use of the free propagator only. The free propagator is the negative of the Feynman propagator, and so the causal and analytic structure at any order in perturbation theory is standard.

The implication of the appearance of complex conjugate poles in the 3+1d ghost theory is still a topic of debate. There has been a recent study \cite{Kubo:2024ysu} of the effect of one-loop matter corrections to a real scalar ghost in 3+1d QFT. These authors emphasize that the complex conjugate poles occur on the physical sheet and they argue that a certain combination of energy eigenstates is a true asymptotic state. This state is the analog of our $\beta$-state, and in their model this is a negative norm state. Our results are suggesting that the construction of a new inner product for use with the Born rule, that is a generalization of our approach to 3+1d QFT, could again yield a probability interpretation.

In \cite{Holdom:2024cfq} we studied a renormalizable 4-derivative scalar field theory, an interacting theory of a massive ghost and a massless normal particle. In the high energy limit it was found that a certain combination emerged as the single asymptotic state. The scattering differential cross section for this state is positive (on or above a critical line in the renormalization group flow diagram). This has some similarities to the situation with the $\beta$-state. This is also the combination of energy eigenstates corresponding to the single asymptotic state, and for which we argue, can yield positive probabilities. 

\section{Numerical results for models with more fields and more spatial points}\label{s10}

We can introduce more than one field into the model Hamiltonian by using the matrix representation of the Kronecker product. For example for two ghost fields $\phi_1$ and $\phi_2$ we have the matrix representation
\begin{equation}
    \phi_1^2+\phi_2^2\to (\bm{\phi}\bm{\eta}\bm{\phi})\otimes\bm{\eta}+\bm{\eta}\otimes(\bm{\phi}\bm{\eta}\bm{\phi})
.\end{equation}
The two different field spaces correspond to the first and second factors of the Kronecker product. The quadratic term in the matrix Hamiltonian $\tilde{\textbf{H}}$ involves the multiplication by $\bm{\eta}\otimes\bm{\eta}$ on the left, which yields
\begin{equation}
    \tilde{\bm{\phi}}^2\otimes1+1\otimes\tilde{\bm{\phi}}^2
.\end{equation}

We shall construct a model with a normal field $\phi_+$ coupled to another field $\phi$ that we take for now to be a ghost, $\sigma=-1$; we compare to the case of $\sigma=1 $ at the end. We would like to see whether the spectrum remains positive when the ghost interacts with a non-ghost. The matrix Hamiltonian has two interaction terms,
\begin{align}
    \tilde{\textbf{H}}&=m_1\bm h\otimes1+m_21\otimes\bm h\nonumber\\&+\frac{\lambda_1}{4m_1\sqrt{m_2}}\bm{\phi}_+^2\otimes\tilde{\bm{\phi}}-\frac{\lambda_2}{(2m_2)^\frac{3}{2}3!}1\otimes\tilde{\bm{\phi}}^3
.\label{e11}\end{align}
We let the first factor of the Kronecker product represent the normal  space, for which we can ignore the tilde. We have chosen both interaction terms to be even in the normal field and odd in the ghost field. This model has two propagators, for the $\phi_+$ and $\phi$ fields respectively, and the required matrix elements are
\begin{align}
    \langle\bar n|\phi_+|\bar m\rangle&=\bm\psi^{(n)\dagger}\left(\bm \phi_+\otimes\bm\eta\right)\bm\psi^{(m)},\\\langle\bar n|\phi|\bar m\rangle&=\bm\psi^{(n)\dagger}\left(1\otimes\bm \phi\right)\bm\psi^{(m)},\\
    \langle\bar n|\bar n\rangle&=\bm\psi^{(n)\dagger}(1\otimes\bm\eta)\bm\psi^{(n)}
.\end{align}

We shall study this model for $m_1=m_2=1$  and as a function of $\lambda_1$ with $\lambda_2=\frac{5}{4}\lambda_1$. The latter choice helps to favor positive energy eigenvalues over complex-conjugate pairs for finite size matrices.  We again obtain positive values of $E_n-E_0$ up to $n=40$ for both weak and strong couplings using a matrix size of $140^2\times140^2$.\footnote{Unlike the model in Section \ref{s4}, the numerical study of this model and the others in this section do not require the extra high precision. Instead the method must deal with larger matrices in a memory efficient way. We use the Krylov method.} The sparseness is $0.0004$. The results are shown in Table \ref{t3}, now for the couplings $\lambda_1=1/3$ and $3$. We only show states that contribute to the ghost propagator through non-vanishing $Z_n$ values, and in fact some eigenstates do not contribute to either propagator. Once again we see that interactions tend to push the energy levels further apart.

\begin{table}[h]
    \centering
    \begin{tabular}{|c|c|c|c|} \hline 
  \multicolumn{2}{|c|}{$\lambda_1=1/3$} & \multicolumn{2}{|c|}{$\lambda_1=3$} \\ \hline 
  $E_n-E_0$& $Z_n$& $E_n-E_0$&$Z_n$\\ \hline 
           0.0&  0.040295606&  0.0 & 0.2796511127\\ \hline 
           1.0654874515&  $-$1.0141543269&  1.9005199208& $-$1.0460010496\\ \hline 
           2.1334468133&   0.0002320954&  3.8557553225& 0.0039481953\\ \hline 
           2.1993651665&   0.0140319684&  4.2480520842&  0.0429192409\\\hline
           3.370413048&   $-$0.0001104&  6.014464618&  $-$6.125e-6\\\hline
    \end{tabular}
    \caption{Ghost$+$non-ghost model in (\ref{e11}) with $\sigma=-1$, $k=3$, $m_1=m_2=1$ and $\lambda_2=\frac{5}{4}\lambda_1$.}
    \label{t3}
\end{table}

The Kronecker product also offers a way to construct models with several spatial points. These models offer some indication of the effect of the spatial derivative part of the kinetic term. The different field spaces of the Kronecker product will now represent the same field at different points in space. The 2-site matrix Hamiltonian is constructed as follows,
\begin{align}
    \tilde{\textbf{H}}&=m(\bm h\otimes1+1\otimes\bm h)\nonumber\\&+\frac{1}{m}\sigma(\tilde{\bm{\phi}}\otimes1-1\otimes\tilde{\bm{\phi}})^2\label{e15}\\&+\frac{\lambda}{(2m)^\frac{k}{2}k!}\left(\tilde{\bm{\phi}}^k\otimes1+1\otimes\tilde{\bm{\phi}}^k\right).\nonumber
\end{align}
$\bm h$ is the matrix defined in (\ref{e3}), and the second line is the spatial derivative term that comes with a factor of $\sigma$.

There are two ways to construct the 2-site model, by using one or two links. The former would imply a $1/2$ factor in the spatial derivative term. We have chosen the latter since it corresponds to the 3-site and 4-site models below, which have three and four points on a circle. With this choice all three models have the number of links matching the number of points.

Once the eigenvalues and eigenvectors are determined from $\tilde{\textbf{H}}$, the matrix elements needed for the calculation of the propagator are given by
\begin{align}
    \langle\bar n|\phi|\bar m\rangle&=\frac{1}{\sqrt{2}}\bm\psi^{(n)\dagger}\left(\bm \phi\otimes\bm\eta+\bm\eta\otimes\bm \phi\right)\bm\psi^{(m)},\\
    \langle\bar n|\bar n\rangle&=\bm\psi^{(n)\dagger}(\bm\eta\otimes\bm\eta)\bm\psi^{(n)}
.\end{align}
We are again able to obtain positive values of $E_n-E_0$ up to $n=40$ for both weak and strong couplings, when using a matrix size of $140^2\times140^2$. The sparseness is $0.0008$. In Table \ref{t5} we again only show the low lying eigenstates that contribute to the propagator.
\begin{table}[h]
    \centering
    \begin{tabular}{|c|c|c|c|} \hline 
  \multicolumn{2}{|c|}{$\lambda=0.1$}& \multicolumn{2}{|c|}{$\lambda=10$}\\ \hline 
  $E_n-E_0$& $Z_n$& $E_n-E_0$&$Z_n$\\ \hline 
           0.0&  0.0006406662&  0.0& 0.228960611\\ \hline 
           1.0012578205&  $-1.0002762122$&  2.2726023342& $-$1.0389359302\\ \hline 
           2.003548823&  0.00027558&  4.8167168499& 0.0362442354\\ \hline 
           &  &  7.3880542435& 0.003108632\\ \hline 
           &  &  7.5431977148& $-$0.0003149398\\\hline
 & & 10.1311397436&$-$0.000106178\\\hline
    \end{tabular}
    \caption{2-site model in (\ref{e15}) with $\sigma=-1$, $k=3$ and $m=1$.}
    \label{t5}
\end{table}

Next we study the 3-site model
 \begin{align}
    \tilde{\textbf{H}}&=m\left(\bm h\otimes1\otimes1+1\otimes\bm h\otimes1+1\otimes1\otimes\bm h\right)\nonumber\\&+\frac{1}{2m}\sigma(\tilde{\bm{\phi}}\otimes1\otimes1-1\otimes\tilde{\bm{\phi}}\otimes1)^2\nonumber\\&+\frac{1}{2m}\sigma(1\otimes\tilde{\bm{\phi}}\otimes1-1\otimes1\otimes\tilde{\bm{\phi}})^2\label{e10}\\&+\frac{1}{2m}\sigma(1\otimes1\otimes\tilde{\bm{\phi}}-\tilde{\bm{\phi}}\otimes1\otimes1)^2\nonumber\\&+\frac{\lambda}{(2m)^\frac{k}{2}k!}\left(\tilde{\bm{\phi}}^k\otimes1\otimes1+1\otimes\tilde{\bm{\phi}}^k\otimes1+1\otimes1\otimes\tilde{\bm{\phi}}^k\right)\nonumber
\end{align}
Here the needed matrix elements are
\begin{align}
    \langle\bar n|\phi|\bar m\rangle&=\frac{1}{\sqrt{3}}\bm\psi^{(n)\dagger}\left(\bm \phi\otimes\bm\eta\otimes\bm\eta+\bm\eta\otimes\bm \phi\otimes\bm\eta+\bm\eta\otimes\bm\eta\otimes\bm \phi\right)\bm\psi^{(m)},\\
    \langle\bar n|\bar n\rangle&=\bm\psi^{(n)\dagger}(\bm\eta\otimes\bm\eta\otimes\bm\eta)\bm\psi^{(n)}
.\label{e13}\end{align}
We obtain positive values of $E_n$ up to $n=40$ for both weak and strong couplings, when using a matrix size of $60^3\times60^3$. The sparseness is $0.0001$. The results are shown in Table \ref{t2}.
\begin{table}[h]
    \centering
    \begin{tabular}{|c|c|c|c|} \hline 
  \multicolumn{2}{|c|}{$\lambda=0.1$}& \multicolumn{2}{|c|}{$\lambda=10$}\\ \hline 
  $E_n-E_0$& $Z_n$& $E_n-E_0$&$Z_n$\\ \hline 
           0.0&  0.0006406662&  0.0& 0.3321954366\\ \hline 
           1.0010302497&  $-$1.0001857341&  2.241382842& $-$1.0324879043\\ \hline 
           2.0027506258&  0.000184028&  4.68190092& 0.0260647897\\ \hline 
           &  &  6.80610809& 0.0068158472\\ \hline 
           &  &  7.26915471& $-$0.0002083844\\\hline
 & & 9.39421413&$-$0.0001535012\\\hline
    \end{tabular}
    \caption{3-site model in (\ref{e10}) with $\sigma=-1$, $k=3$ and $m=1$.}
    \label{t2}
\end{table}

Lastly we study the 4-site model. Its specification is an obvious extension of the last model, and so we do not give the analogs of (\ref{e10}-\ref{e13}). We obtain positive values of $E_n$ up to $n=40$ for both weak and strong couplings, when using a matrix size of $27^4\times27^4$. The sparseness is $0.00006$. As the number of sites increases the kinetic terms leave many more degeneracies, some of which are not broken by the interactions. The numerical procedure is not guaranteed to find all the degenerate eigenstates. The results for the eigenstates that are found to contribute to the $Z_n$ are shown in Table \ref{t6}. Comparing the results for the 2, 3, and 4-site models, we find that they are quite similar, especially for the lower lying eigenstates. The main finding is that positive spectra survive the effects of the spatial derivative terms in the multi-site models.
\begin{table}[h]
    \centering
    \begin{tabular}{|c|c|c|c|} \hline 
  \multicolumn{2}{|c|}{$\lambda=1/10$}& \multicolumn{2}{|c|}{$\lambda=10$}\\ \hline 
  $E_n-E_0$& $Z_n$& $E_n-E_0$&$Z_n$\\ \hline 
           0.0&  0.00077366&  0.0& 0.4400718949\\ \hline 
           1.0009379538&  $-$1.0001415431&  2.2329862423& $-$1.0296124881\\ \hline 
           2.0023939548&  0.0001381203&  4.6199108185& 0.0193460987\\ \hline 
           &  &  6.141718683& 0.0087146052\\ \hline 
           &  &  7.12540144& $-$0.0001580782\\\hline
 & & 7.30163186&0.0020010723\\\hline
    \end{tabular}
    \caption{4-site model with $\sigma=-1$, $k=3$ and $m=1$.}
    \label{t6}
\end{table}

Now we extend the two field model to two spatial points, so that we combine the two different kinds of fields and their interactions with the effects of a spatial derivative term for each field. We restrict to a common mass $m$. The first two factors of the Kronecker product represent the two fields at one site and the last two represent the two fields at the other site.
 \begin{align}
    \tilde{\textbf{H}}&=m\left(\bm h\otimes1\otimes1\otimes1+1\otimes\bm h\otimes1\otimes1\right.\nonumber\\&+\left.1\otimes1\otimes\bm h\otimes1+1\otimes1\otimes1\otimes\bm h\right)\nonumber\\&+\frac{1}{m}(\bm{\phi}_+\otimes1\otimes1\otimes1-1\otimes1\otimes\bm{\phi}_+\otimes1)^2\nonumber\\&+\frac{1}{m}\sigma(1\otimes\tilde{\bm{\phi}}\otimes1\otimes1-1\otimes1\otimes1\otimes\tilde{\bm{\phi}})^2\label{e33}\\&+\frac{\lambda_1}{4m^\frac{3}{2}}\left(\bm{\phi}^2_+\otimes\tilde{\bm{\phi}}\otimes1\otimes1+1\otimes1\otimes\bm{\phi}^2_+\otimes\tilde{\bm{\phi}}\right)\nonumber\\&-\frac{\lambda_2}{(2m)^\frac{3}{2}3!}\left(1\otimes\tilde{\bm{\phi}}^3\otimes1\otimes1+1\otimes1\otimes1\otimes\tilde{\bm{\phi}}^3\right)\nonumber
\end{align}
The required matrix elements are
\begin{align}
    \langle\bar n|\phi_+|\bar m\rangle&=\frac{1}{\sqrt{2}}\bm\psi^{(n)\dagger}\left(\bm \phi_+\otimes\bm\eta\otimes 1\otimes\bm\eta+\bm 1\otimes\bm\eta\otimes\bm \phi_+\otimes\bm\eta\right)\bm\psi^{(m)},\\\langle\bar n|\phi|\bar m\rangle&=\frac{1}{\sqrt{2}}\bm\psi^{(n)\dagger}\left(1\otimes\bm \phi\otimes1\otimes\bm \eta+1\otimes\bm \eta\otimes1\otimes\bm \phi\right)\bm\psi^{(m)},\\
    \langle\bar n|\bar n\rangle&=\bm\psi^{(n)\dagger}(1\otimes\bm\eta\otimes1\otimes\bm\eta)\bm\psi^{(n)}
.\end{align}
We again choose $\lambda_2=\frac{5}{4}\lambda_1$, and we again have positive values of $E_n-E_0$ up to $n=40$ for both weak and strong couplings using a matrix size of $27^4\times27^4$. The sparseness is $0.00005$. The results are shown in Table \ref{t4}. A positive spectrum survives both the interactions between ghosts and non-ghosts as well as the spatial derivative terms.

\begin{table}[h]
    \centering
    \begin{tabular}{|c|c|c|c|} \hline 
  \multicolumn{2}{|c|}{$\lambda_1=1/3$} & \multicolumn{2}{|c|}{$\lambda_1=3$} \\ \hline 
 $E_n-E_0$ & $Z_n^-$ & $E_n-E_0$ & $Z_n^-$\\ \hline 
           0.0& 0.0388994588&  0.0 & 0.4248130637\\ \hline 
           1.0426732387& $-$1.0082971592&  1.7822157097& $-$1.0352257155\\ \hline 
           2.0871155592&  0.0001599798&  3.6091153986& 0.0037406963\\ \hline 
           2.1259847379&  0.0081512058&  3.9057586883&  0.0306135004\\\hline
           3.232460172&  $-$4.0674e-5&  5.569916703&  $-$6.317e-6\\\hline
    \end{tabular}
    \caption{Ghost$+$non-ghost 2-site model in (\ref{e33}) with $\sigma=-1$, $k=3$, $m=1$ and $\lambda_2=\frac{5}{4}\lambda_1$.}
    \label{t4}
\end{table}

In this section we have considered five different models involving a ghost field. In each model we see that the interactions cause the energy levels to be pushed further apart, and this provides some understanding as to why complex conjugate states do not occur. In addition, in each model the ghost can be turned into a non-ghost by setting $\sigma=1$. But then each model is subject to the usual consequences of cubic interactions. For each truncated model with $\sigma=1$ we find that there is a coupling $\lambda$ above which the spectrum is no longer bounded from below. This shows up by having negative values of $E_n-E_0$ interspersed among the positive ones, when ordered by absolute value. The existence of this phenomenon indicates that the normal theories ($\sigma=1$) do not exist, at least for strong coupling. In contrast, we have provided evidence that the ghost theories ($\sigma=-1$) do exist for both weak and strong coupling.

\section*{Acknowledgements}
I thank James Stokes for working with me during the early stage of this work. I thank Niayesh Afshordi, Luca Buoninfante, Simon Caron-Huot, Jisuke Kubo, Taichiro Kugo, Alberto Salvio and Alessandro Strumia for discussions.

\end{document}